\documentclass[twocolumn,secnumarabic,amssymb, nobibnotes, aps, prd]{revtex4-1}

\usepackage{xspace}%
\usepackage[normalem]{ulem} 
\usepackage{soul}

\newcommand{\newoutG}[1]{%
\xspace
}

\newcommand{\newout}[1]{%
\xspace
}
\usepackage{upgreek} 

\usepackage{epsfig, amsmath,amsfonts, amssymb,graphicx,amsthm,color}

\usepackage{mathtools}
\usepackage{dcolumn}
\usepackage{bm}
\usepackage{rotating}
\usepackage{amssymb}
\usepackage[latin1]{inputenc}
\usepackage{comment}
\usepackage{multirow}
\usepackage{tabularx}
\usepackage[pagebackref=false]{hyperref}

\hypersetup{
    colorlinks=true,       
    linkcolor=blue,          
    citecolor=blue,       
    filecolor=magenta,      
    urlcolor=blue          
}
\newcommand {\Ket}[1]         {\ensuremath{| \, #1 \rangle}} 
 
\graphicspath{{Images/}} 
\usepackage{lipsum}
\usepackage[symbol]{footmisc}

\usepackage{filecontents}
{

\begin{document}

\title{Searching optimal conditions for quantum gates application with the new 3-body F\"orster resonances in Rb and Cs Rydberg atoms}


\author{K.-L. Pham, S. Lepoutre, P. Pillet and P. Cheinet}
\email[Contact: ]{patrick.cheinet@u-psud.fr}
\affiliation{Laboratoire Aim\'{e} Cotton, CNRS, Univ. Paris-Sud, Universit\'e Paris-Saclay, B\^{a}t. 505, 91405 Orsay, France }
\author{I. N. Ashkarin, I. I. Beterov, D. B. Tretyakov, V. M. Entin, E. A. Yakshina and I. I. Ryabtsev}
\email[Contact: ]{ryabtsev@isp.nsc.ru}
\affiliation{Rzhanov institute of semiconductor physics, Pr. Lavrentyeva 13, 630090 Novosibirsk, Russia}

\date{\today}

\begin{abstract}
\noindent\textbf{Abstract} Three body resonant interactions between Rydberg atoms are considered in order to perform few-body quantum gates. So far, the resonances found in cesium or rubidium atoms relied on an adjacent two-body resonance which ceases to exist for principal quantum numbers above $n \simeq 40$. We have proposed recently a new class of 3-body interaction resonances in alkali-metal Rydberg atoms [P. Cheinet \textit{et al.}, Quant. Elect. \textbf{50}, 213 (2020)], which circumvienes this limit. We investigate here the relative strength between this new class of 3-body interaction resonance and quasi-forbidden 2-body interaction resonances in rubidium and cesium Rydberg atoms. We then identify the best case scenario for detecting and using this 3-body interaction.
\end{abstract}

\maketitle

\section{Introduction}

Rydberg atoms \cite{GAL} offer an ideal platform for quantum simulation experiments \cite{WEI10} or quantum computing \cite{SAF10}. The high control gained on ultra-cold atoms trapping \cite{BAR16,END16} enabled new physics studies with Rydberg atoms \cite{DEL19,BRO20,MAD20} involving the particularly large interactions displayed. 

In this prospect, interaction resonances tuned with electric field \cite{SAFGAL81} have been extensively studied in alkali-metals \cite{GAL82,STO87,AND98,MOU98,VOG06,REI08,RYA10,ALT11,NIP12,TRE14,RAV15} with a recent attention to their coherence properties~\cite{RAV14,YAK16}.

Stark tuned F\"ortser resonances have recently been revisited towards few-body resonances \cite{GUR12,FAO15,TRE17,LIU20} and their use for quantum computation is now under investigation with theoretical studies on their coherence \cite{RYA18} and on specific Toffoli gate protocol~\cite{BET18}. 

In these investigations, the 3-body resonance presented below in equation (\ref{Eq_Forster_np3B}) benefits from the nearby 2-body F\"orster resonance presented below in equation (\ref{Eq_Forster_np2B}) to lead to a strong interaction. Considering rubidium or cesium atoms in a given principal quantum number $n$, an orbital quantum number $\ell=1$ often referred to as $p$ and a total momentum quantum number $j=3/2$, these resonances can be written:
\begin{equation}
\left|3 \times np_{3/2}\right\rangle\Leftrightarrow\left|ns_{1/2}+(n+1)s_{1/2}+np'_{3/2}\right\rangle
\label{Eq_Forster_np3B}
\end{equation}
\begin{equation}
\left|2 \times np_{3/2}\right\rangle\Leftrightarrow\left|ns_{1/2}+(n+1)s_{1/2}\right\rangle
\label{Eq_Forster_np2B}
\end{equation}

with the $\left|np'_{3/2}\right\rangle$ state corresponding to a Rydberg atom in the same level but a different Zeeman sub-level. The 2-body resonances of type (\ref{Eq_Forster_np2B}) exist in rubidium and cesium atoms but are limited to principal quantum numbers of around $n\simeq40$ at most and, as a consequence, the related 3-body resonances stop as well. But such interaction resonances gain in strength with increasing $n$ and obtaining a 3-body resonance for unlimited $n$ is thus of high interest. 

One possibility to overcome the limit is to use atoms in different initial states as proposed in \cite{BET18}. But it then requires the use of several phase-locked lasers at different frequencies which represents an additional technical difficulty that may prevent a successfull implementation. More recently, a new class of 3-body resonance was proposed \cite{Cheinet_2020} to overcome the $n$ limit while still using all atoms in a single initial state:
\begin{equation}
\left|3 \times np_{3/2}\right\rangle\Leftrightarrow\left|ns_{1/2}+(n+1)s_{1/2}+np_{1/2}\right\rangle.
\label{Eq_Forster_np3B12}
\end{equation}

In this new scenario, the use of $\left|np_{1/2}\right\rangle$ state allows to accomodate for larger intermediate state detunings in the 3-body resonance such that it survives the disappearence of the 2-body resonance. But it also implies a weaker 3-body interaction by typically an order of magnitude. In a previous study \cite{PEL16}, we have demonstrated that quasi-forbidden resonances can have non negligible strength for principal quantum numbers of around $n\simeq30$ compared to the 2-body interaction. 

In this article, we thus study the relative strength between the new class of 3-body resonances and quasi-forbidden resonances nearby in rubidium or cesium Rydberg atoms. Strong quasi-forbidden resonances will indeed perturb gate protocols and may reduce significantly the resulting gate fidelity. This study allows us to conclude on optimum scenario to perform a quantum gate from this new resonant interaction scheme.

\section{ \label{Sec2} Theoretical methods}
\begin{figure}[t]
	\includegraphics[width=9cm]{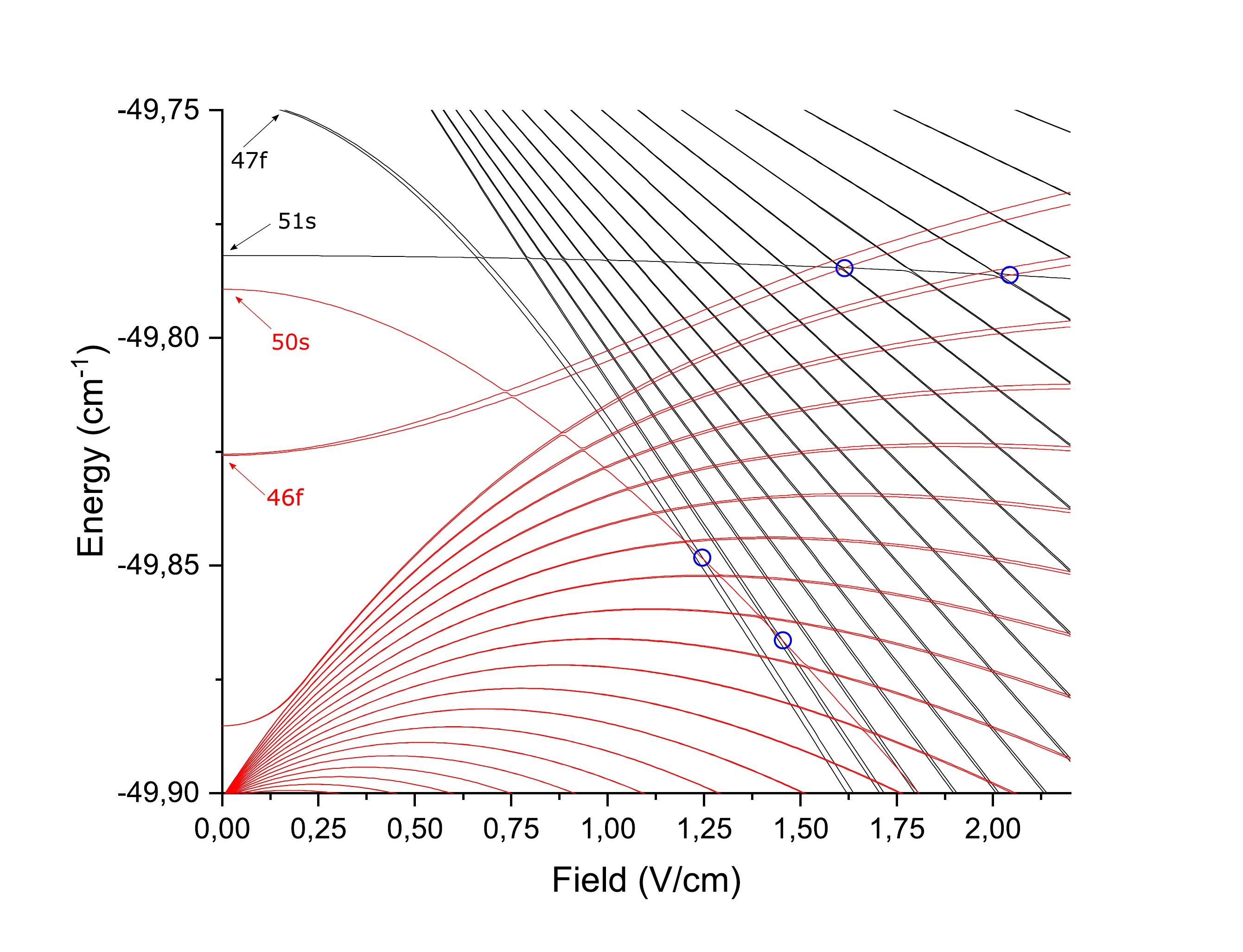}
	\caption{\label{Stark} Stark F\"orster Resonance diagram from $\left|50p_{3/2}\right\rangle$ starting state of cesium at an energy of around -50.9 cm$^{-1}$, limited to $|m|=1/2$ states. (black) Stark energies $E_{nlj}$. (red) Symmetric energy $2E_{50p3/2}-E_{nlj}$. The crossings between these curves are possible F\"orster resonances. The blue circles show the resonances which follow the condition $\delta_\ell=\pm1$ for at least one state at zero field and for which we calculate interaction strength.}
\end{figure}

For this study, we follow the methodology already presented in \cite{PEL16} that we remind briefly. We start by computing the Stark energies of all states around the considered $\left|np_{3/2}\right\rangle$ starting state of a given 3-body resonance of type (\ref{Eq_Forster_np3B12}). In this study, we restricted our calculations to states $|m|=1/2$ to avoid searching through too numerous resonances. It was demonstrated in \cite{PEL16} that resonances involving higher $|m|$ display similar strengths. We then obtain a resonance map, as displayed from $\left|50p_{3/2}\right\rangle$ of cesium in figure \ref{Stark}: we plot in black the Stark energies $E_{nlj}$ of all states in a given energy interval above the starting state and in red we plot the symmetric energy $2 E_{50p_{3/2}} - E_{nlj}$. Any crossing between a black line and a red line then corresponds to a possible interaction resonance. The strongest are the ones for which at least one state follows the zero field selection rule $\delta_\ell=\pm1$ \cite{PEL16}. We have marqued in figure \ref{Stark} the first resonances implying states $\left|50s_{1/2}\right\rangle$ or $\left|51s_{1/2}\right\rangle$. For the electric field range, we arbitrarily choose to search up to around twice the 3-body resonance field.

In our search through different energy ranges in cesium, we find the following resonances:
\begin{align}
\left|2 \times np_{3/2}\right\rangle\Leftrightarrow& \left|ns_{1/2}+(n-3)l_{j}\right\rangle \label{Eq_Forster_Quasi1}\\
\left|2 \times np_{3/2}\right\rangle\Leftrightarrow& \left|(n+1)s_{1/2}+(n-4)l_{j}\right\rangle \label{Eq_Forster_Quasi2}\\
\left|2 \times np_{3/2}\right\rangle\Leftrightarrow& \left|(n+1)p_{1/2}+(n-2)d_{j}\right\rangle \label{Eq_Forster_Quasi3}\\
\left|2 \times np_{3/2}\right\rangle\Leftrightarrow& \left|(n-1)p_{1/2}+nd_{j}\right\rangle \label{Eq_Forster_Quasi4}\\
\left|2 \times np_{3/2}\right\rangle\Leftrightarrow& \left|(n+1)d_{j}+(n-3)d_{j}\right\rangle. \label{Eq_Forster_Quasi5}
\end{align}

When the resonances have been identified, we compute the corresponding dipole-dipole interaction energy using the transition dipole moments between the two starting $\left|np_{3/2}\right\rangle$ states and the two ending Stark states \cite{PEL16}. They will depend, among other things, on the overlap integral between the radial wavefunctions which is large for states close in energy to the starting state. 

The closest states leading to resonances and fulfilling the $\delta_\ell=\pm1$ selection rule are the two states $\left|ns_{1/2}\right\rangle$ and $\left|(n+1)s_{1/2}\right\rangle$. The first two resonances (\ref{Eq_Forster_Quasi1}) and (\ref{Eq_Forster_Quasi2}) are thus the strongest as found previously \cite{PEL16}. We also note that the resonance (\ref{Eq_Forster_Quasi5}) is fully allowed regarding the selection rules but the involved states are separated from the initial state by about 2 multiplicities leading to small overlap integrals.

Due to significantly different quantum defects in rubidium Rydberg atoms, the resonances found for this atom are of different type:
\begin{align}
\left|2 \times np_{3/2}\right\rangle\Leftrightarrow& \left|(n-1)s_{1/2}+(n-1)l_{j}\right\rangle \label{Eq_Forster_Quasi6}\\
\left|2 \times np_{3/2}\right\rangle\Leftrightarrow& \left|(n+2)s_{1/2}+(n-4)l_{j}\right\rangle \label{Eq_Forster_Quasi7}\\
\left|2 \times np_{3/2}\right\rangle\Leftrightarrow& \left|(n-2)d_{5/2}+(n-2)l_{j}\right\rangle \label{Eq_Forster_Quasi8}\\
\left|2 \times np_{3/2}\right\rangle\Leftrightarrow& \left|(n-1)d_{5/2}+(n-3)l_{j}\right\rangle. \label{Eq_Forster_Quasi9}
\end{align}

We note that in rubidium, the strong quasi-forbidden resonances of type (\ref{Eq_Forster_Quasi1}) and (\ref{Eq_Forster_Quasi2}) do not appear any more. They occur for fields of typically more than twice the 3-body resonance field. The resonances involved are all of quasi-forbidden type with large energy difference with the initial state, implying small overlap integrals in the dipole-dipole interaction.

In the resonances (\ref{Eq_Forster_Quasi1}-\ref{Eq_Forster_Quasi9}) listed above, the generic label $l$ indicates that the resonance occurs with states of the Stark multiplicity leading to a large set of consecutive resonances. We do not compute the whole set but the first ones in order to grasp their general behavior. The generic label $j$ is used when the two possible values $j=l-1/2$ and $j=l+1/2$ lead to resonances. For the sets of resonances with a multiplicity, we compute one out of the two as they were found of similar strength \cite{PEL16}.

\section{ \label{Sec3} Results}
\label{Results}
\begin{figure}[t]
	\includegraphics[width=10cm]{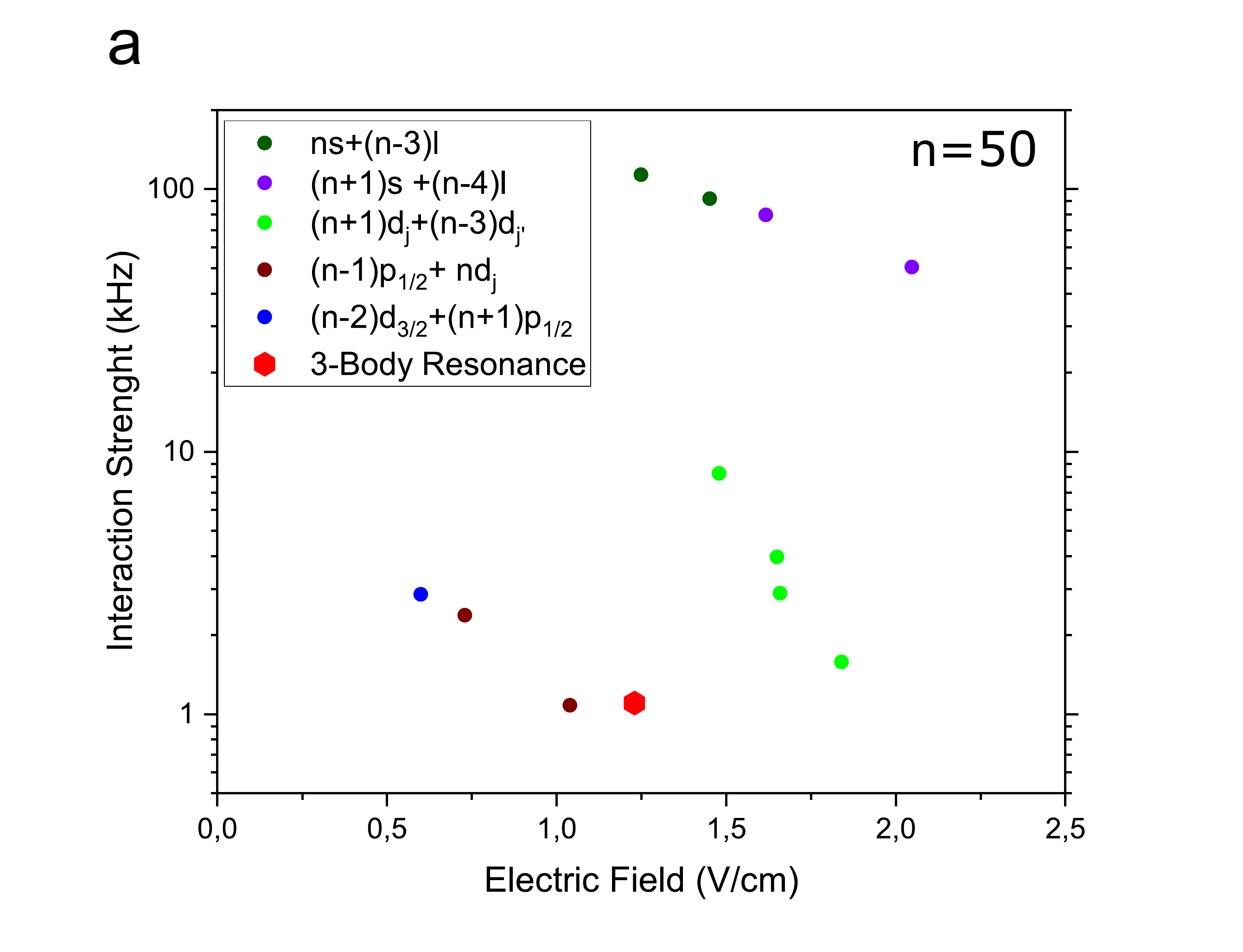}
	\includegraphics[width=10cm]{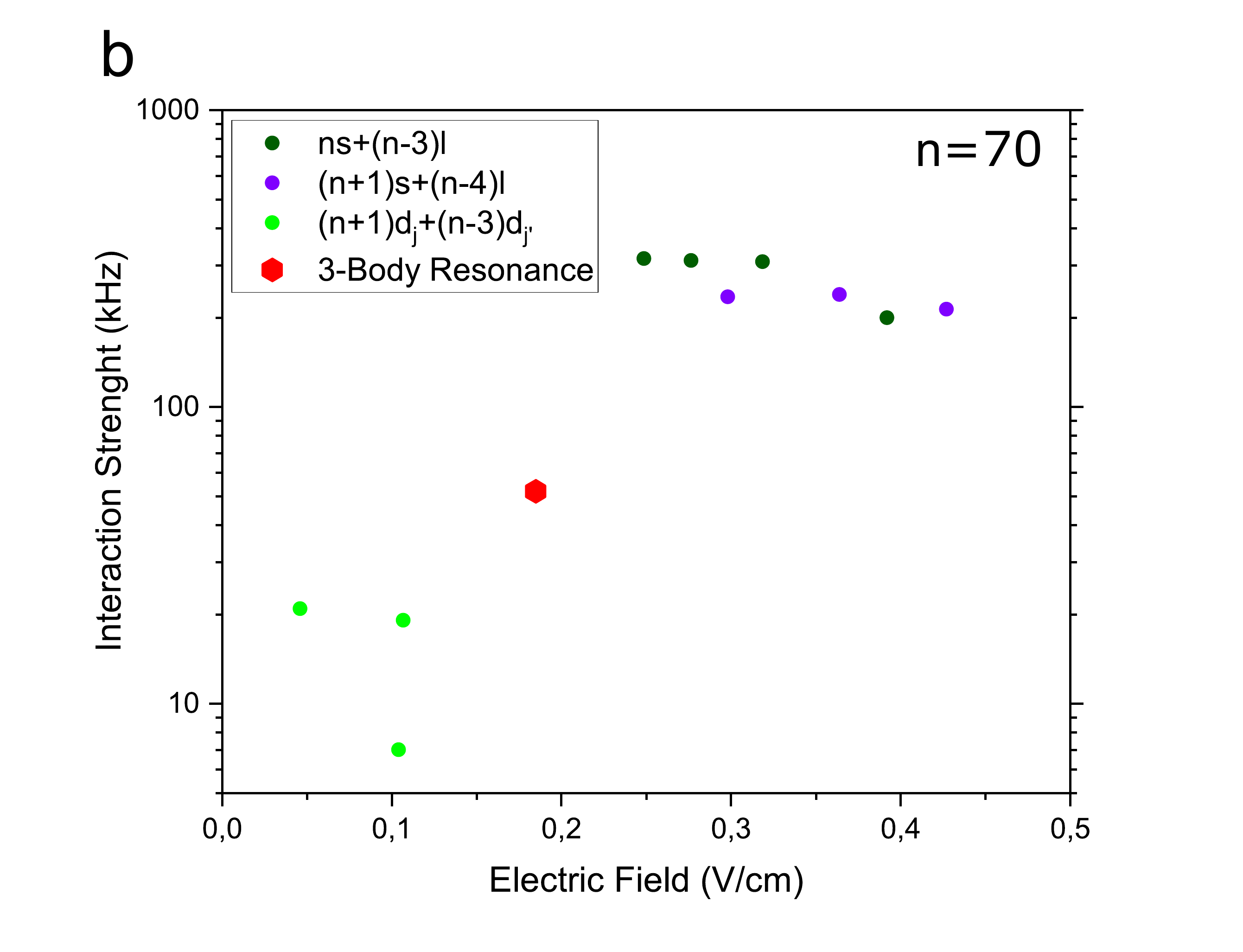}
	\caption{Interaction strength at 10 $\mu m$ interatomic distance versus Electric Field for various resonances in Cesium Rydberg atoms from initial state $\left|np_{3/2}\right\rangle$. (a) Resonances at $n=50$. The 3-body resonance is close to much stronger 2-body resonances. (b) resonances at $n=70$. The 3-body resonance is still weaker than many close 2-body resonances. }
	\label{cesium}
\end{figure}

\begin{figure}[t]
	\includegraphics[width=10cm]{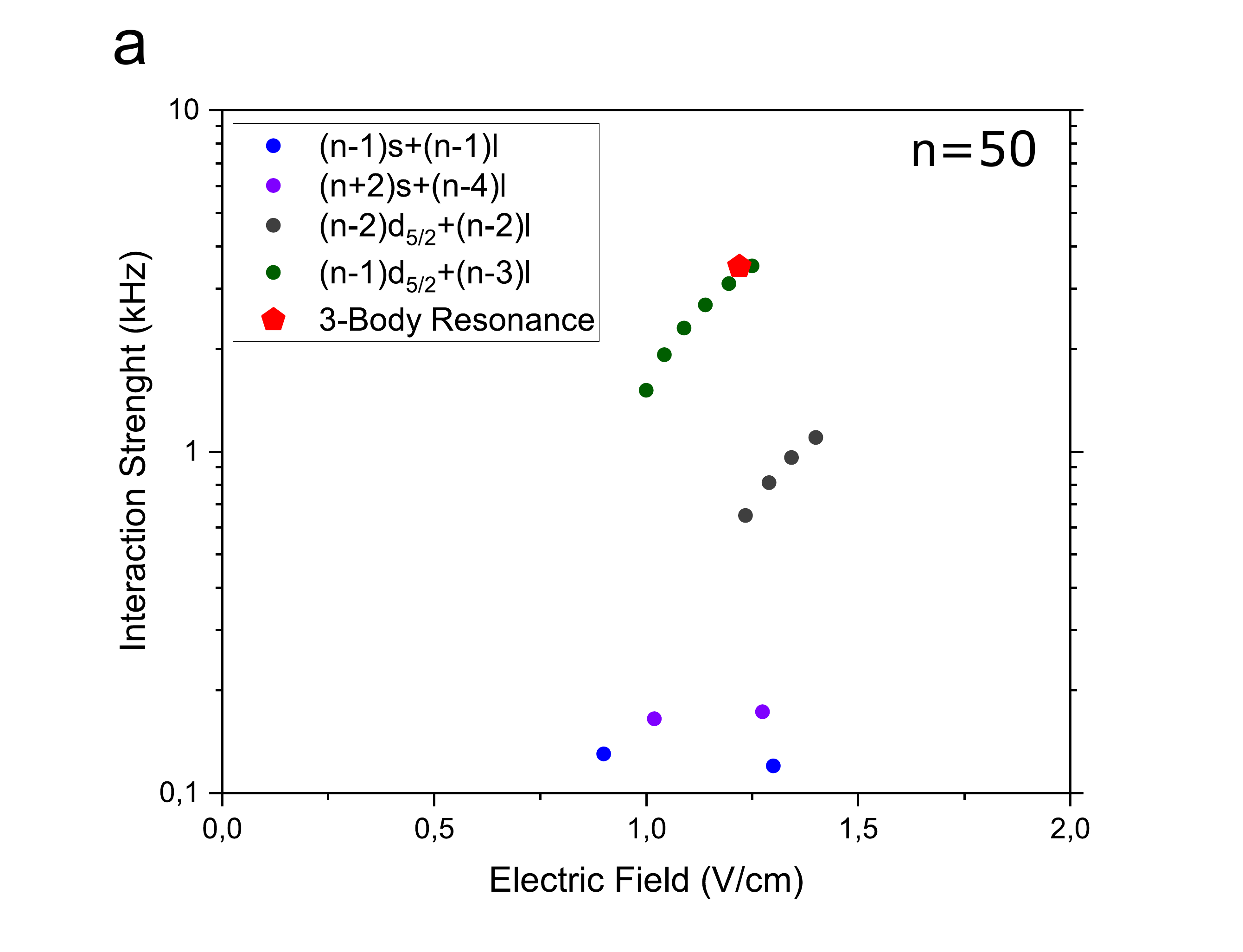}
	\includegraphics[width=10cm]{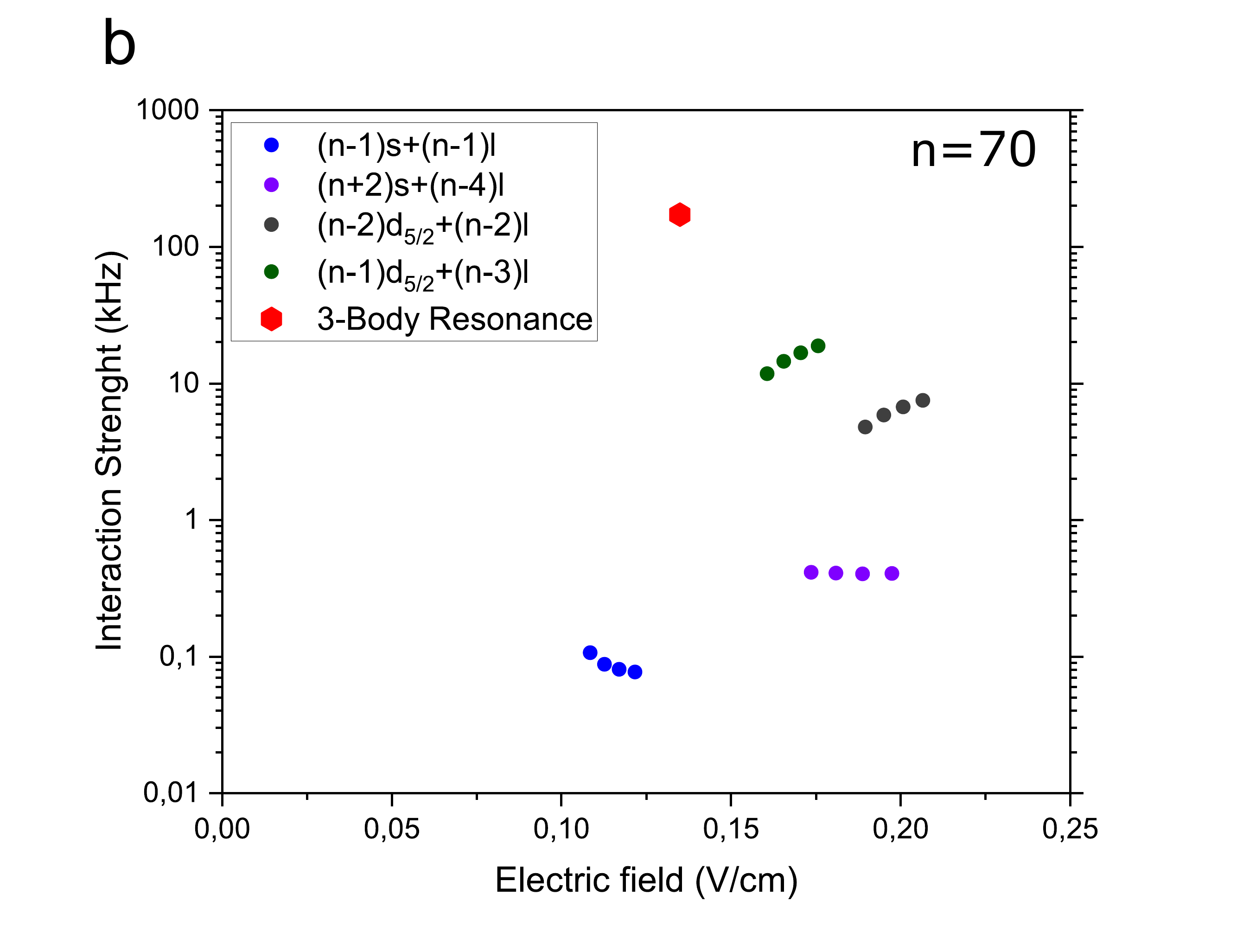}
	\caption{Interaction strength at 10 $\mu m$ interatomic distance versus Electric Field for various resonances in Rubidium Rydberg atoms from initial state $\left|np_{3/2}\right\rangle$. (a) Resonances at $n=50$. The 3-body resonance is embedded with numerous 2-body resonances of similar strengths. (b) resonances at $n=70$. The 3-body resonance is now stronger and well seperated from all surrounding 2-body resonances.}
	\label{rubidium}
\end{figure}

F\"orster resonances have already been studied on cesium \cite{MOU98,VOG06,NIP12} and rubidium \cite{AND98,WES06,RYA10} atoms below $n\simeq$ 40 where the 2-body resonance (\ref{Eq_Forster_np2B}) exists, together with the 3-body resonance (\ref{Eq_Forster_np3B}). Therefore we concentrate on principal quantum numbers above the limit and choose $n=50$ and $n=70$ to test the evolution with increasing $n$. We present the results of the different resonances in cesium and rubidium atoms in figures \ref{cesium} and \ref{rubidium} which display their interaction strength and compare to the corresponding 3-body resonance of type (\ref{Eq_Forster_np3B12}). As we consider a potential application to quantum computation and the interaction strength is highly dependent on distance, we compute the interaction strength for a relatively large interatomic distance of 10 $\mu m$ which is likely to be well controlled experimentally. All interaction strengths displayed in figures (\ref{cesium}) and (\ref{rubidium}) are also gathered in tables (\ref{cesiumT}) and (\ref{rubidiumT}).

In cesium atoms, in figure (\ref{cesium}), we can see that at $n=50$ the 3-body resonance is embedded within numerous quasi-forbidden resonances which are up to a hundred times stronger. It is therefore unlikely to achieve any reliable quantum computation with this interaction scheme at this principal quantum number $n$. Increasing $n$ will increase all interaction strengths but the 3-body one, which scales as $n^{11}$ will increase faster than the 2-body ones which scale as $n^4$. We thus check at $n=70$ and see that the 3-body interaction strength has indeed significantly increased comparatively to the 2-body ones. It also stands almost alone at its resonance field of 0.19 V/cm. Nevertheless, it is still almost an order of magnitude weaker than the next quasi-forbidden resonances of types (\ref{Eq_Forster_Quasi1}) and (\ref{Eq_Forster_Quasi2}). Increasing $n$ further will keep strengthening the 3-body interaction compared to the quasi-forbidden resonances. However the sensitivity to electric field increases as well, as $n^7$, and will eventually become an issue. Another option is to use a smaller interatomic distance. Indeed the 3-body interaction scales as $1/R^6$ while the 2-body interactions scale as $1/R^3$ but the sensitivity to the exact distance will then increase and might as well become an issue in a gate protocol fidelity. It is thus interesting to check wether rubidium atoms present better caracteristics.

In rubidium Rydberg atoms, the energy difference between $\Ket{np_{3/2}}$ and $\Ket{np_{1/2}}$ is smaller than in cesium. This implies that the consecutive 3-body resonance is stronger. We also explained that the resonances (\ref{Eq_Forster_Quasi1}) and (\ref{Eq_Forster_Quasi2}) occur for significantly higher fields than the 3-body resonance. Thus they should not impair a gate protocol. Instead, we found resonances (\ref{Eq_Forster_Quasi6}-\ref{Eq_Forster_Quasi9}) surrounding the 3-body resonance with significant strength. In figure \ref{rubidium}, we present the strengths of these resonances together with the 3-body resonance, at $n=50$ and $n=70$. We can see that resonances of type (\ref{Eq_Forster_Quasi9}) are the strongest and are, at $n=50$, of similar strength and resonant field as the 3-body resonance. Therefore it seems again unlikely to find suitable experimental parameters at this principal quantum number to perform a high fidelity 3-body quantum gate. Instead, at $n=70$, the 3-body resonance has become the strongest interaction by almost an order of magnitude while, like in cesium atoms, its resonance field becomes seperate from the resonance field of the quasi-forbidden processes. 

For increasing $n$, the relative strength of the 3-body resonance will keep increasing compared to the spurious quasi-forbidden resonances and will completely dominate them even at interatomic distances of up to 10 microns. We thus conclude that rubidium atoms are indeed suited to implement 3-body quantum gates based on this interaction resonance. 


\begin{table}[t!]
	\begin{center}
		\begin{tabular}{c|c|c|c|c|}
			\cline{2-5}
			&\multicolumn{2}{c}{n=50}& \multicolumn{2}{|c|}{n=70}\\
			\cline{2-5}
			&F(V/cm) & I.S.(kHz) &F(V/cm) & I.S.(kHz)\\
			\cline{1-5}
			\multicolumn{1}{ |c|}{ns+(n-3)l}& 1.25 &113.568 &0.25&316.617\\
			\cline{2-5}
			\multicolumn{1}{ |c|}{} & 1.45& 91.953 & 0.28&311.760\\
			\cline{2-5}
			\multicolumn{1}{ |c|}{} & & & 0.32&308.427\\
			\cline{2-5}
			\multicolumn{1}{ |c|}{}& & & 0.39& 199.902\\
			\hline\hline
			\multicolumn{1}{ |c|}{(n+1)s+(n-4)l} & 1.62 & 79.728 &0.30&234.933\\
			\cline{2-5}
			\multicolumn{1}{|c|}{}& 2.05&50.453 & 0.36&239.299 \\
			\cline{2-5}
			\multicolumn{1}{|c|}{}& & &0.43&213.391\\
			\cline{2-5}
			\multicolumn{1}{|c|}{}&& &&\\
			\hline\hline
			\multicolumn{1}{|c|}{(n+1)d$_{5/2}$+(n-3)d$_{5/2}$} &1.48& 8.250& & \\
			\cline{1-5}
			\multicolumn{1}{|c|}{(n+1)d$_{3/2}$+(n-3)d$_{5/2}$} &1.65 &3.970 & 0.05&20.874 \\
			\cline{1-5}
			\multicolumn{1}{|c|}{(n+1)d$_{5/2}$+(n-3)d$_{3/2}$} &1.66 &2.890 & 0.11&19.118 \\
			\cline{1-5}
			\multicolumn{1}{|c|}{(n+1)d$_{3/2}$+(n-3)d$_{3/2}$} &1.84 &1.580 &0.10 &6.990 \\
			\cline{1-5}
			\hline\hline
			\multicolumn{1}{|c|}{(n-1)p$_{1/2}$+nd$_{5/2}$} &0.73 &2.380 & & \\
			\cline{1-5}
			\multicolumn{1}{|c|}{(n-1)p$_{1/2}$+nd$_{3/2}$} &1.04 &1.080 & & \\
			\cline{1-5}
			\hline\hline
			\multicolumn{1}{|c|}{(n-2)d$_{3/2}$+(n+1)p$_{1/2}$} &0.60 &2.860 & & \\
			\hline\hline
			\multicolumn{1}{|c|}{ns+(n+1)s+np$_{1/2}$}&1.23 &1.100 &0.19 &52.0\\
			\hline
		\end{tabular}
		\caption{Cesium computed resonance strengths. For each quasi-forbidden resonance of types (\ref{Eq_Forster_Quasi1} - \ref{Eq_Forster_Quasi5}), the resonance electric field F and expected interaction strength (I.S.) is reported. It is compared at the end of the table with the 3-body resonance.}
		\label{cesiumT}
	\end{center}
\end{table}

\begin{table}[t]
	\begin{center}
		\begin{tabular}{c|c|c|c|c|}
			\cline{2-5}
			&\multicolumn{2}{c}{n=50}& \multicolumn{2}{|c|}{n=70}\\
			\cline{2-5}
			&F(V/cm) & I.S.(kHz) &F(V/cm) & I.S.(kHz)\\
			\cline{1-5}
			\multicolumn{1}{ |c|}{(n-1)s+(n-1)l}& 0.90 & 0.130&0.109&0.106\\
			\cline{2-5}
			\multicolumn{1}{ |c|}{} &1.30 &0.120  &0.113&0.088\\
			\cline{2-5}
			\multicolumn{1}{ |c|}{} & & &0.117 &0.081\\
			\cline{2-5}
			\multicolumn{1}{ |c|}{}& & &0.122 &0.077 \\
			\hline\hline
			\multicolumn{1}{ |c|}{(n+2)s+(n-4)l} & 1.02 &0.165  &0.17&0.413\\
			\cline{2-5}
			\multicolumn{1}{|c|}{}&1.28&0.173 &0.18& 0.406\\
			\cline{2-5}
			\multicolumn{1}{|c|}{}& & &0.19&0.403\\
			\cline{2-5}
			\multicolumn{1}{|c|}{}&& &0.20&0.404\\
			\hline\hline
			\multicolumn{1}{|c|}{(n-2)d$_{5/2}$+(n-2)l} &1.24 &0.650 &0.190 &4.794 \\
			\cline{2-5}
			\multicolumn{1}{|c|}{} &1.29&0.810 &0.195 &5.856 \\
			\cline{2-5}
			\multicolumn{1}{|c|}{} &1.34 &0.960 &0.201 &6.721 \\
			\cline{2-5}
			\multicolumn{1}{|c|}{} &1.4&1.100 &0.207 &7.513 \\
			\hline\hline
			\multicolumn{1}{|c|}{(n-1)d$_{5/2}$+(n-3)l} &1.00&1.510 &0.161 &11.766 \\
			\cline{2-5}
			\multicolumn{1}{|c|}{} & 1.04& 1.920& 0.166& 14.459\\
			\cline{2-5}
			\multicolumn{1}{|c|}{}&1.09&2.300&0.171&16.710\\
			\cline{2-5}
			\multicolumn{1}{|c|}{}&1.14&2.690&0.176&18.821\\
			\cline{2-5}
			\multicolumn{1}{|c|}{}&1.20&3.100&&\\
			\cline{2-5}
			\multicolumn{1}{|c|}{}&1.25&3.500&&\\
			\hline \hline
			\multicolumn{1}{|c|}{ns+(n+1)s+np$_{1/2}$}&1.22 &3.493 &0.14 &172\\
			\hline
		\end{tabular}
		\caption{Rubidium computed resonance strengths. For each quasi-forbidden resonance of types (\ref{Eq_Forster_Quasi6} - \ref{Eq_Forster_Quasi9}), the resonance electric field F and expected interaction strength (I.S.) is reported. It is compared at the end of the table with the 3-body resonance.}
		\label{rubidiumT}
	\end{center}
\end{table}

\section{Conclusion} 
In summary, we have investigated the quasi-forbidden resonances surrounding the 3-body interaction resonance scheme (\ref{Eq_Forster_np3B12}) presented in \cite{Cheinet_2020} and evaluated their relative strengths both in cesium and rubidium Rydberg atoms. This study points out that rubidium is better suited for implementing quantum calculations with 3-body interactions than cesium, provided the use of a high enough principal quantum number. Future studies should concentrate on $n=70$ or higher in rubidium atoms. The next step towards quantum calculations is to evaluate specific gate protocols functionning with this interaction scheme. Expected key parameters to obtain good fidelities are a high degree of control of atomic positions and of the electric field to achieve high coherence times.

\section{Acknowledgements}
The french team was supported by the EU H2020 FET Proactive project RySQ (grant N. 640378). The russian team was supported by Novosibirsk State University. This work was also supported by cooperation grant ECOMBI (CNRS grant No PRC2312 and RFBR grant No 19-52-15010).

\bibliographystyle{apsrev4-1}

\bibliography{biblio_3B_P12}

\end{document}